# HERIOT-WATT UNIVERSITY

# SFA Referee Allocation Scheme

**Nikolaos Polatidis**

| BSc Computer Science Honours Dissertation | |
|---|---|
| Title | SFA Referee Allocation Scheme |
| Author | Nikolaos Polatidis |
| Matric No | 033351337 |
| Supervisor | Phil Trinder |
| Second reader | Mike Chantler |
| Date | 02 June 2008 |

# Declaration

I, Nikolaos Polatidis, confirm that this work submitted for assessment is my own and is expressed in my own words. Any uses made within it of the works of other form (e.g., ideas, equations, figures, text, tables, programs) are properly acknowledged at any point of their use. A list of references employed is included.

Signed:

Date:

# Abstract

For many sports, the allocation of officials to matches is performed manually and is a very time consuming procedure. For the Scottish Football Association (SFA), the allocation of referees and other officials to matches is governed by a number of rules specifying the expertise required from the different types of official at each level, e.g. Scottish Premiership League referee must be a grade 1 with high experience. The allocation requires an SFA secretary to expend several hours to find suitable officials, contact them and assign them. Most of the time, the secretary is a volunteer who performs the allocation as a hobby and it would be useful to reduce his costs and time.

The project aims to reduce the burden on SFA, and potentially other secretaries, by developing a program to assign SFA officials. A suitable algorithm must be devised to search through the set of data about matches and officials and find a potential allocation. The program then updates the database with the new data, and provides a web interface for both secretaries and officials.

The project was proposed by the author and its achievements of can be summarized as follows. To prepare the background for the project, a literature review of referee allocation has been undertaken, together with an investigation of existing allocation tools, and consultation with the SFA. The requirements for a prototype allocator were established as well as possible by consultation with the SFA. A range of algorithms and tools for solving the referee allocation problem have been investigated, including backtracking/constraint-solving approach. A new greedy allocation algorithm has been developed.

A prototype system using the new greedy algorithm has been implemented and evaluated with SFA secretaries. A final usable referee allocation system has been designed that uses the greedy algorithm, and is extended after evaluation of the prototype. The final allocation system based provides both a command line and a web interface and has also been evaluated by SFA secretaries. In their letters of recommendation in Appendix F the SFA secretaries indicate that the final allocation system it will be used again in the future.

# Acknowledgements

I would like to thank Dr. Phil Trinder for his immeasurable help and support throughout this project. I would also like to thank Mike Chantler for his constructive feedback. I also feel the need to express my thanks to Iain McCrone for his valuable help.

Finally, I would like to acknowledge the constant support offered by my parents.

# Table of Contents

# List of Figures

# List of Tables

# Chapter 1

# Introduction

This chapter covers an introduction to the project including the context, a description of aims and objectives, a description of what has been achieved, a description of the methodology used and an outline.

## 1.1 Context

Referee allocation is a very time consuming procedure. For the Scottish Football Association (SFA), the allocation of referees and other officials to matches is governed by a number of rules specifying the expertise required from the different types of official at each level, e.g. Scottish Premiership League referee must have grade 1 experience. Section 2.1 contains a longer explanation of the allocation problem and section 2.3 contains a discussion of referee allocation tools currently available. Although the allocation works properly as it is, it takes more than a few hours to find the suitable officials, contact them and assign them. This process could be simplified and improved with the use of computer software. Most of the times the secretary is a volunteer who does this as a hobby, therefore it would be useful to reduce his costs and time.

The approach proposed at this project shows potential for the future since it will be able to assign referees, assistant referees, fourth officials and observers in a very short period of time and reduce costs. The system gives the ability to the league secretary, through a suitable interface, to add, edit or delete teams, fixtures or officials, assign officials both manually and automatically for a certain date, change an assignment and view the assignments. Also the web interface gives the ability to the officials to state their availability for a certain date and view the assignments.

## 1.2 Aims and Objectives

The aim of this project is to assign officials to different levels of football fixtures either automatically, using a greedy algorithm, or manually.

The main objectives for the project are as follows:

**Database:** To develop a database were all the relevant information about the officials; the fixtures and the assignments will be stored.

**Allocator:** To develop an algorithm for SFA official allocation, together with a program to perform the allocation at different levels of football.

**Website:** To develop a Web Interface which aims to give the ability to the officials to state their availability, using a username and a password, for a certain date. Among this the option to view the assignments will be available.

## 1.3 Achievements

The following goals were achieved:

- Investigated a range of algorithms for solving the referee allocation problem, including backtracking/constraint solving and developed a greedy allocation algorithm (section 4.1).
- Designed a usable referee allocation system, based on the greedy algorithm, which

has been extended after consultation with the SFA (section 4.1).

- Implemented a prototype system and evaluated it with SFA secretaries. Implemented a final system, based on the prototype, with both a command line and a web interface (section 5.2).

- Evaluated the final system with SFA secretaries. The system has been used by SFA secretaries to assign officials and it will be used again in the future (section 6.2).

- Completed a literature research within the SFA to learn about the referee allocation procedure in general, established the requirements with consultation with the SFA and investigated existing tools (section 2.1).

## 1.4 Methodology

The design and development of the system followed the prototyping methodology. Prototyping is a vital part of software development.

"A prototype can be anything from a paper-based storyboard through to a complex piece of software, and from a cardboard mockup to a molded or pressed piece of metal [1]."

A high fidelity software based prototype was used to allow potential users to interact with the product, validate and negotiate the requirements. The prototype was functional, had the look and feel of the final product and its design concepts were evaluated by potential users.

## 1.5 Outline of Dissertation

This dissertation is split into several chapters.

- Chapter two is the background research, which gives an overview of the referee assignment, presents the logical rules governing the assignments and gives an examples assignment. It also mentions the programming language used, the initial plan for the solving the problem and the web application technologies.

- Chapter three covers the requirements for the Allocator. Is explains the role of the system and presents the requirements for the Command Line system and the Web Interface. It also describes the functionality of the final system.

- Chapter four covers the design and evaluation of the prototype system. It includes the initial database design, an explanation of the initial algorithm and the new developed algorithm. It goes on to explain the functionality provided by both the Command Line system and the web interface. There is also a section on how it was evaluated. The algorithm used can be found in section 4.1.2.2.

- Chapter five covers the implementation of the final system. It includes system diagrams explaining the functionality. The final system is also described with further detail on its operation.

- Chapter six contains the testing and evaluation of the final system.

- Chapter seven is the conclusion of the project. It contains a summary, the main achievements of the system, its limitations and future work.

# Chapter 2

# Background Research

This chapter contains the background research about topics relevant to the referee assignment. It presents an overview of the referee allocation and a brief outline of the referee assignment in section 2.1. Any relevant tools available are mentioned at section 2.2. A backtracking approach was initially explored. However for efficiency and simplicity reasons a greedy algorithm was developed and described in section 2.3. In section 2.3 the programming language used and the requirements that it must satisfy can be found. The web application technologies used are available in section 2.4.

## 2.1 Referee Assignment Overview

Referee assignment is the process of assigning officials to a fixture. This can be done by matching referee and assistant referee suitability for the fixture according to rules specified. The two main rules are the class of the referee and the experience, followed by availability, the number of assignments of a referee per day and the number of assignments of a referee for a particular team over a certain period of time. There are certain stages that need to be satisfied in order to provide a usable referee allocation software.

For example in Scotland a category 1 referee is qualified to referee any level of match. All category 2 and category 3 referees are qualified to officiate as assistant referees at any level of match and officiate at semi professional leagues. A category 4 referee is qualified to officiate at semi professional leagues and assist at any matches of the same type. A Category 5 referee and below is qualified to referee at amateur and youth football and be assistant referee up to semi professional leagues.

No referee, assistant referee or fourth official referee is permitted to officiate at more than one game per day.

The Professional leagues are the Scottish Premier League, Scottish Football league one, Scottish Football league 2 and Scottish Football League 3. The semi professional league is called Junior Football League. There are also the amateur leagues and the youth leagues from U11 up to U21.

A fourth official is required only at Premier League matches. Scottish football league one, two, three and the junior league require three officials. Amateur and youth games require one official.

All Premier League matches have observers whereas to other matches people go to observe the officials only when the circumstances allow it.

Every premier league observer must have high experience whereas any SFL one, two, or three could have either high or medium. For any other type of match any type of experience will be sufficient.

### 2.1.1 Logic rules governing assignment

The Following logic rules formally specify much of the referee allocation problem:

(1)

For every Scottish premier league match the category of the referee must be one and his experience high. The category of the assistant referees must be greater than or equal to three. The experience of the observer must be high.

$Premier(match) \Rightarrow Level(referee) = 1 \wedge Experience(referee) = High \wedge Level(assref1) \geq 3 \wedge Level(assref2) \geq 3 \wedge Level(fourthofficial) = 1 \wedge Experience(observer) = High$

(2)

For every Scottish football league one match the category of the referee must be one and his experience either high or medium. The category of the assistant referees must be greater than or equal to three. The experience of the observer could be either high or medium.

$SFL1(match) \Rightarrow Level(referee) = 1 \wedge (Experience = High \vee Experience = Medium) \wedge Level(assref1) \geq \wedge Level(assref2) \geq 3 \wedge (Experience(observer) = High \vee Experience(observer) = Medium)$

(3)

For every Scottish football league two match, the category of the referee must be one and his experience high or medium. The category of the assistant referees must be greater than or equal to three. The experience of the observer could be either high or medium.

$SFL2(match) \Rightarrow Level(referee) = 1 \wedge (Experience = High \vee Experience = Medium) \wedge Level(assref1) \geq 3 \wedge Level(assref2) \geq 3 \wedge (Experience(observer) = High \vee Experience(observer) = Medium)$

(4)

For every Scottish football league three match, the category of the referee must be one and his experience high, medium or low. The category of the assistant referees must be greater than or equal to three. The experience of the observer could be either high or medium.

$SFL3(match) \Rightarrow Level(referee) = 1 \wedge (Experience = High \vee Experience = Medium \vee Experience = Low) \wedge Level(assref1) \geq 3 \wedge Level(assref2) \geq 3 \wedge (Experience(observer) = High \vee Experience(observer) = Medium)$

(5)

For every junior football league match, the category of the referee must greater than or equal to four. The category of the assistant referees must be greater than or equal to six. The experience of the observer could be high, medium or low.

$Junior(match) \Rightarrow Level(referee) \geq 4 \wedge Level(assref1) \geq 6 \wedge Level(assref2) \geq 6 \wedge (Experience(observer) = High \vee Experience(observer) = Medium \vee Experience(observer) = Low)$

(6)

For every mateur football league match, the category of the referee must greater than or equal to six. The experience of the observer could be high, medium or low.

$Amateur(match) \Rightarrow Level(referee) \geq 6 \wedge (Experience(observer) = High \vee Experience(observer) = Medium \vee Experience(observer) = Low)$

(7)

For every youth football league match, the category of the referee must greater than or equal to six. The experience of the observer could be high, medium or low.

$Youth(match) \Rightarrow Level(referee) \geq 6 \wedge (Experience(observer) =High \vee Experience(observer) = Medium \vee Experience(observer) = Low)$

(8)

Every official who officiated at a match at a particular day can not officiate at another match the same day.

$\forall o\ E Official,\ M,M' E\ Match,\ d E Date.\ (o,m,d) \Rightarrow \neg(o,m',d)$

### 2.1.2 Example Assignment

Tables 2.1 and 2.2 list the referees and with their categories and the observers with their experience.

| Referee | Category | Id |
|---|---|---|
| Crawford Allan | 1 | R001 |
| Douglas McDonald | 1 | R002 |
| Thomas Robertson | 1 | R003 |
| Michael Tumilty | 1 | R004 |
| John Underhill | 1 | R005 |
| Brian Colvin | 2 | R006 |
| Paul Reid | 2 | R007 |
| James Bee | 3 | R008 |
| Graham Chambers | 3 | R009 |
| Michal Jasinski | 3 | R010 |
| Keith Sorbie | 3 | R011 |
| Michael Banks | 4 | R012 |
| Gavin Duncan | 4 | R013 |
| Gavin Ross | 4 | R014 |
| Simon MacLean | 5 | R015 |
| Nikolaos Polatidis | 5 | R016 |
| Alastair Wright | 5 | R017 |
| Ewan Young | 5 | R018 |
| Scott Townsley | 6 | R019 |
| Robert Wilson | 6 | R020 |

**Table 2.1** A list of registered officials

| Observer | Experience |
|---|---|
| George Clyde | High |
| George Smith | High |
| Kenny Hope | High |
| David McCartney | High |
| Peter Peace | Medium |
| Neil McLennan | Low |

**Table 2.2** A list of registered observers

| Type | Location | Date | Time | Team 1 | Team 2 | Fixture id |
|---|---|---|---|---|---|---|
| SPL | Edinburgh | 01/12/2007 | 14:00 | Hearts | Hibernian | SPL001 |
| SPL | Aberdeen | 01/12/2007 | 14:00 | Aberdeen | Celtic | SPL002 |
| SFL 1 | Dundee | 01/12/2007 | 14:00 | Dundee | Livingston | SLF001 |
| Junior | Whitburn | 01/12/2007 | 14:30 | Whitburn | Corssgates | J001 |
| Junior | Bonnyrigg | 01/12/2007 | 14:30 | Bonnyrigg | Longside | J002 |
| Youth U21 | Duddingston | 01/12/2007 | 13:30 | Cavalry Park | Liberton United | Y001 |
| Youth U19 | Easthouses | 01/12/2007 | 13:30 | Star | Bonnyrigg | Y002 |

**Table 2.3** A list of fixtures

In order to describe the referee assignment problem better table 2.4 below lists the fixtures covered.

| Fixture id | Referee | Assistant Referee 1 | Assistant Referee 2 | Fourth Official | Observer |
|---|---|---|---|---|---|
| SPL001 | Douglas McDonald | Brian Colvin | James Bee | Thomas Robertson | George Smith |
| SPL002 | Crawford Allan | Graham Chambers | Michal Jasinski | Mike Tumilty | Kenny Hope |
| SLF001 | John Underhill | Keith Sorbie | Paul Reid | | Peter Peace |
| J001 | Michael Banks | Ewan Young | Nikolaos Polatidis | | Neil McLennan |
| J002 | Gavin Ross | Simon MacLean | Alastair Wright | | |
| Y001 | Robert Wilson | | | | George Clyde |
| Y002 | Scott Townsley | | | | David McCartney |

**Table 2.4** An assignment of officials to matches

Above we can see an example with fixtures required officials. According to the rules specified by the SFA the officials and the observers were assigned to the matches.

At the moment referee allocation is a very time consuming operation since any time officials or observers are required the league secretary searches through its handbook for hours to find suitable officials and observers to cover fixtures

## 2.2 Referee Assignment Tools Available

The Referee Assistant software developed by Jeff Wigal [2] assigns referees to games by using a simple ms access database along with a website. The Soccer Central software developed by Shana Saavedra and Janet Irigoyen [3] informs us that it has many features related with referee allocation but it wasn't possible for me to test it due to its pricing.

The tools mentioned above have been developed and maintained in the United States of America. In Scotland, like every other country in Europe, football has some key differences than soccer in the US, one of them being the type of the game. For example one of the previous mentioned tools that I managed to use informs us that the available types of games are High school, Indoor, Intramural and Recreation , whereas in Scotland we have four professional leagues, one semi professional, amateur and youth leagues. Therefore the tools mentioned previously are not suitable enough to be used in Scotland or anywhere in Europe, due to the differences in the types of the games. Furthermore no freeware tool is available at the moment and for financial reasons the Scottish Football Association wants a free utility.

## 2.3 Programming Languages & Libraries

### 2.3.1 Language Selection

A general programming language must be used which has to satisfy certain requirements. It is essential to be able to communicate with the database, it should give the ability to provide a graphical user interface and it must be free. Java has been selected because it satisfies all the previous requirements [4].

A test took place, which involved a connection between Java and MySQL. Basic transfers of data in both directions were implemented successfully.

### 2.3.2 Constraint Solving Tools

The initial selection was a constraint satisfaction problem. A number of libraries that can be used to solve such a problem were explored. SWI-Prolog was tested and works with Java with the use of the JPL package. The communication works both ways. Descriptions of these technologies can be found in Appendix E.

## 2.4 Web Applications

The classic architecture is the three tier architecture as described in Figure 2.1 [5]. The advantages of using this type of model is that the functionality is distributed across three independent systems and also that any of the tier can be replaced or upgraded independently if the requirements or the technology changes.

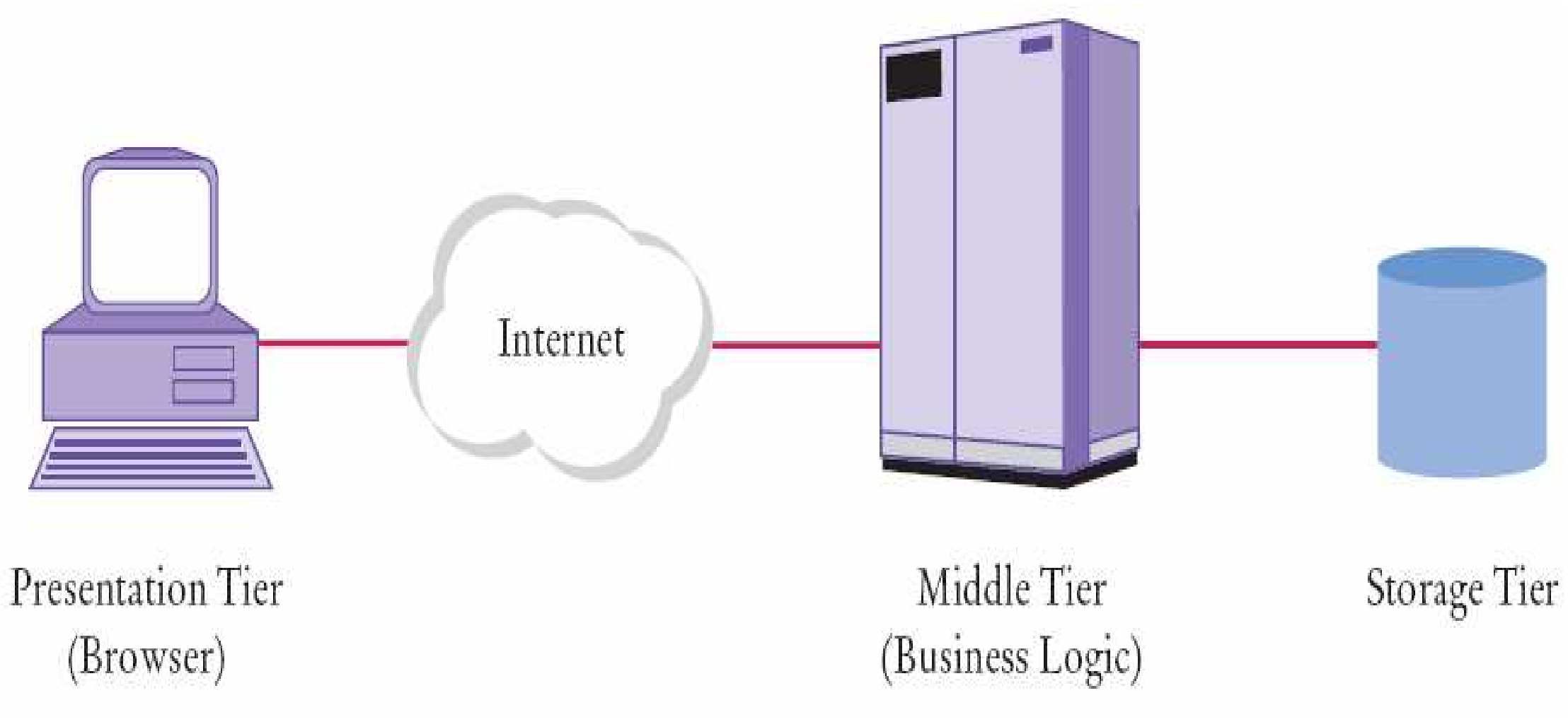

**Figure 2.1** Three Tier Architecture

### 2.4.1 Web applications technologies

There is a number of Database management systems that can be used for this purpose e.g. Oracle, Microsoft SQL server and MySQL ; However MySQL is wise choice since it comes under the General public license and can be used freely. MySQL supports multi threading, multiple user. MySQL is very famous for being easy to work with many programming languages including PHP for creation of dynamic web sites [6]. There is a number of scripting languages e.g. ASP, JSP and PHP. However I think that PHP should be used because it can be embedded into HTML and can run in a web browser and it comes under a GPL compatible license. It can use objects since it is object oriented and it can run under any OS. The way it works is as follows, first it takes the code as input and then it creates web pages as output [7].

PHP was tested with MySQL and it works. Basic tests included transfers to and from the database.

# Chapter 3

# Requirements

This chapter contains the updated version of the requirements specification of the system. Section 3.1 describes the role of the system. The requirements for the Allocator are described in section 3.2 and the web interface requirements in section 3.3, both of which are prioritized. The requirements regarding the optional statistics part of the project are described in section 3.4. The use case model can be found in section 3.5.

## 3.1 Role of the System

The role of the initial system is:

> To assign referees, assistant referees, fourth officials and observers to football games.

The role of the final system is:

- To provide a solver that will automatically and manually assign officials to football games, provide functionality for multiple weeks, change assignments, pre assign officials and view assignments.
- To provide a web interface that the officials will use to add information using a username and a password and display assignment information regarding the fixtures that they are assigned to, without any validation.

- To provide a web interface that will display statistical information about the officials. This part is optional.

## 3.2 Command Line System Requirements

The rules that were described at section 2.1.1 will be applied.

The secretary that will have access to the Allocator, will also be able to perform any of the following operations using a suitable interface:

- Add and delete fixtures
- Add and delete referees
- Assign officials
- Change assignments
- Pre assign officials
- View assignments

### 3.2.1 Referee allocation

[Priority High]

The task is to develop software that will either automatically allocate officials to football fixtures. The algorithm will be based on assignment rules that need to be satisfied in order to assign the officials.

### 3.2.2 Pre assignment

[Priority High]

The task is to provide the functionality to assign a number of officials manually and the rest automatically.

### 3.2.3 Multiple weeks

[Priority High]

The task is to give the ability to the secretary to assign officials for a particular date and therefore for multiple weeks.

### 3.2.4 Selective assignment

[Priority High]

The ability to assign officials for a certain league, such as SPL, SFL1, SFL2, SFL3 and juniors will be available. Among this the options to assign officials for all leagues for a specified date will be available.

### 3.2.5 Editing an assignment

[Priority High]

The option to edit an assignment will be available. The option aims to give the option to remove an official from an assignment and allocate another at his place. Then the removed official will be available for any future assignments for the particular date.

## 3.3 Web Interface

### 3.3.1 Official availability

[Priority High]

A web site will be implemented where the officials will be able to state their availability for a certain date. Every official will have a unique user name and password which will use to gain access to the site.

### 3.3.2 Official allocation

[Priority High]

A website displaying the appointments will be available.

### 3.3.3 Observer

[Priority Low]

Every observer will have access to a site by using a user name and a password where they will be able to add, delete or edit referee marks. The marks will then change the referee's experience if appropriate.

### 3.3.4 Clubs

[Priority Medium]

Every club will have access to a web site by entering a user name and a password. The site will display information about the fixture. Among the information that will be displayed will be the names of the officials, the name of the teams and the date.

### 3.3.5 Stadiums

[Priority Medium]

A certified official from every stadium will be able to check the fixtures that will be played at the particular stadium. Again a user name and a password will be necessary to gain access to the site.

## 3.4 Statistics

[Optional Part]

A web site which will display statistical information about the officials will be implemented. The following information will be provided:

- **Correctness of Decision:** Application and Interpretation of the Laws of the Game.

  Statistical information about the selected referee's ability to apply and interpret correctly the laws of the game will be displayed.

- **Match Control:** Tactical Approach and Management of the Game.

  Statistical information about the selected referee's ability to be in control of the game will be displayed here. For example if a player simulates a foul and the referee gives the foul other's will then see it and he or she will, probably lose control. The ability to be in control of the game using common sense will be measured statistically.

- **Management of Players and Team Officials:** Disciplinary Control

This part includes the referee's ability to manage the players and the team officials in a way that they won't give dissent and respect the official.

- **Average number of sanctions**

  The average number of yellow and red cards that a referee uses at every fixture will be measured statistically.

All the previous information will be based on data retrieved from the database. The data will be based on the report marks submitted by the observers. The information will adjust accordingly every time a new report is submitted by any of the observers.

## 3.5 Use case model

The use case diagram describes the functionality of the final system as designed from the requirements and can be found below in figure 3.1.

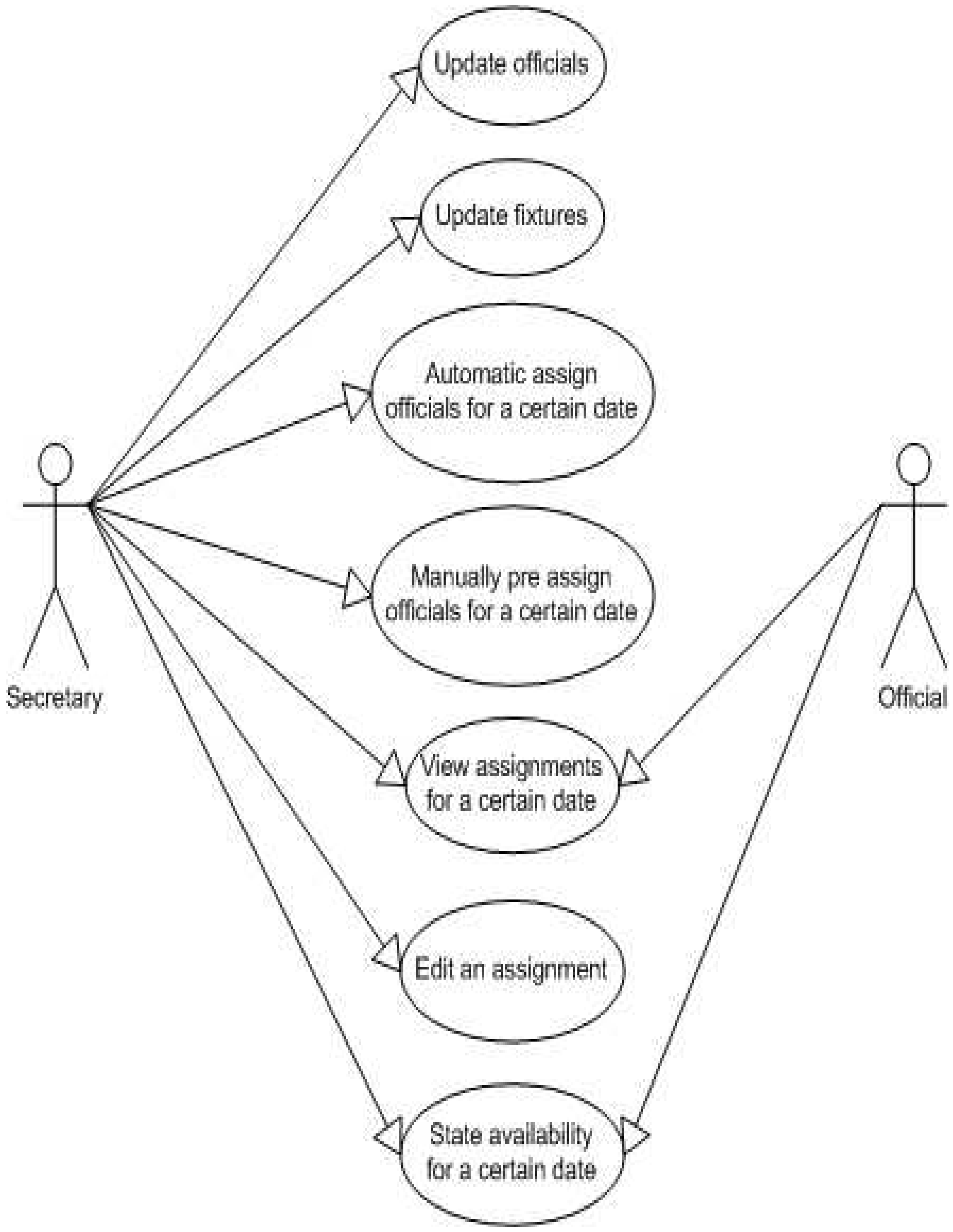

**Figure 3.1** UML Use Case diagram

# Chapter 4

# Prototype Design and Evaluation

This chapter outlines the design and evaluation of the prototype, which is very similar to the final system as described in section 5.1. The prototype is functional, usable and includes both a user interface and a web site. The database and the Allocator design can be found in section 4.1 and the evaluation in section 4.2. The algorithm used can be found in section 4.1.2.2. The methodology used for the development of the system can be found in section 1.4.

## 4.1 Design

### 4.1.1 Database Design

There are a number of Database management systems that can be used for this purpose e.g. Oracle, Microsoft SQL server and MySQL. However MySQL is a wise choice since it comes under the Gnu and can be used freely. MySQL supports multi threading and multiple users. MySQL is very famous for being easy to work with many programming languages including PHP for creation of dynamic web sites.

Figure 4.1 below shows the Entity Relationship diagram designed for the database of the prototype.

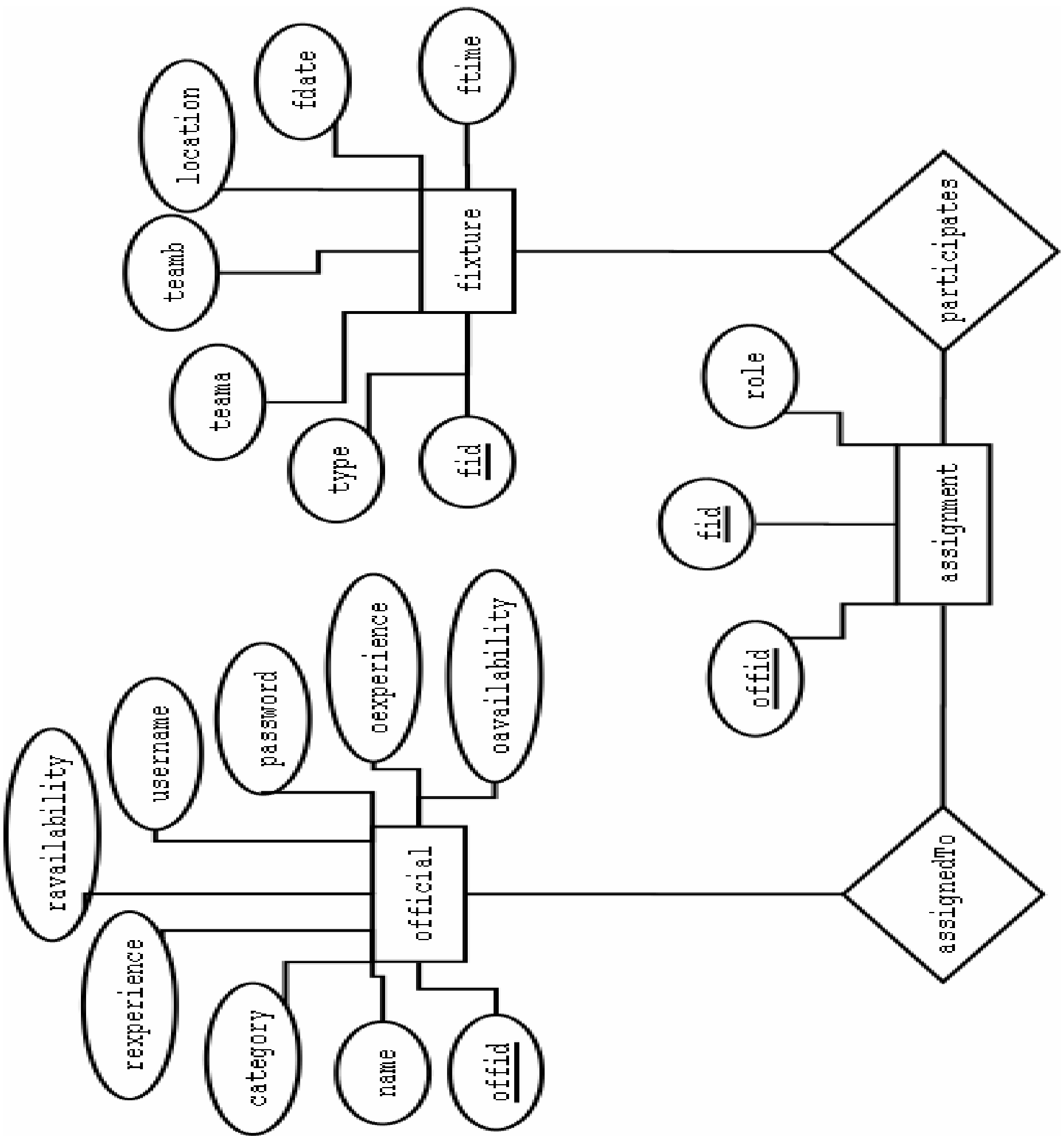

**Figure 4.1** SFA Prototype Allocator Entity Relationship Diagram

### 4.1.2 Allocation Algorithms

#### 4.1.2.1 Backtracking

Backtracking is a type of algorithm that uses brute force search to find all possible solutions. A number of solutions might be rejected if they do not fulfill certain requirements. Backtracking can be used to find potential solution to constraint satisfaction problems [8].

Consider the following referee allocation example in order to understand better how the backtracking algorithm works.

There are two fixtures that need from two officials each and there are four officials available. Table 4.1 gives the name and the grade of the official.

| Name | Category |
|---|---|
| A | 1 |
| B | 1 |
| C | 2 |
| D | 2 |

**Table 4.1** A number of officials

The two fixtures that need two officials each are M1 and M2. Both M1 and M2 need a category 1 and a category 2 or greater referee respectively.

The algorithm will then proceed as follows:

| | | |
|---|---|---|
| M1 | A | B |
| M2 | …. | |

**Table 4.2** An attempt to assign

Referee A and B will be assigned to M1 but M2 won't be satisfied, therefore the algorithm will backtrack and produce something like the following:

| | | |
|---|---|---|
| M1 | A | C |
| M2 | B | D |

**Table 4.3** An assignment of officials to matches

#### 4.1.2.2 The Greedy Method

"In the greedy method we attempt to construct an optimal solution in stages. At each stage we make a decision that appears to be the best at the time (under some criterion) at the time. A decision made in one stage is not changed in a later stage, so each decision should assure feasibility. The criterion used to make the greedy decision at each stage is called the greedy criterion" [9].

Consider the following referee allocation example in order to understand better how the greedy algorithm works.

There are two fixtures M1 and M2 that need three officials each.

M1 needs a category 1 official and two category 2 officials.

M2 needs a category 1 official and two category 3 officials.

However for M2 there is an extra criterion stating that the two category 3 officials could be of a superior category up to 2.

Table 4.5 lists the available officials:

| Name | Category |
|---|---|
| A | 1 |
| B | 1 |
| C | 2 |
| D | 2 |
| E | 2 |
| F | 3 |
| G | 3 |

**Table 4.4** A number of officials

Then the algorithm will produce the following:

| | | | |
|---|---|---|---|
| M1 | A | C | D |
| M2 | B | E | F |

**Table 4.5** An assignment of officials to matches

The category 1 officials were assigned as required and the category 2 officials were assigned as required. However in the M2 case the second official is a category 2 and not 3 since the larger value is assigned first.

A major issue that can arise is assigning the officials with the highest category first and then the ones with a lower category and so on, thus leaving most of the time the officials of a small category without a game and category one officials with more games. With the use of a greedy algorithm and the logic rules mentioned at section 2.1.1 this would be the case. However in practice this is not possible because the SFA wants every official of a certain category to participate more at they highest level possible for them so they can gain experience and proceed. This can be solved by making the algorithm stricter.

Consider the following pseudo code example to see how the actual algorithm will assign SPL officials:

```
SELECT officials IF category is one AND referee experience is high THEN
        SELECT fixtures IF type is SPL THEN
                Assign officials as referees FOR the number of fixtures
SELECT officials IF category is two OR category is three THEN
        Assign officials as assistant referees 1 FOR the number of fixtures
SELECT officials IF category is two OR category is three THEN
        Assign officials as assistant referees 2 FOR the number of fixtures
SELECT officials IF category is one AND experience is high THEN
        Assign officials as fourth officials FOR the number of fixtures
SELECT officials IF observer experience is high THEN
        Assign officials as observers FOR the number of fixtures
```

**Figure 4.2** Pseudo code SPL Example

## 4.2 Evaluation

The following units of the Allocator were evaluated by two SFA secretaries, the discussion of which can be found at section 4.3.

### 4.2.1 Update officials and fixtures

The update officials and fixtures option of the menu takes us to a new menu with the options to add or remove officials and fixtures and return to the main menu.

### 4.2.2 Assign officials

The assign officials, option takes us to a new menu with the options to assign officials at Scottish premier league, Scottish football league 1, Scottish football league 2, Scottish football league 3 and junior football. The assignment can be done separately for every league or with the assign all option which will assign officials to every fixture.

### 4.2.3 View assignments

The view appointments option gives us the opportunity to display on the screen the appointments to fixtures. This can be done for every league separately or for all leagues.

### 4.2.4 Change assignment

The option to change an assignment is given.

### 4.2.5 Example assignment

The command line trace below display how the assignment would appear if the user decided to view the assignments using the view option at the Allocator. Also I would like to mention that the view assignments option displays the same assignments as the screenshot in figure 4.4. The assignment tables refer only to SPL fixtures.

```
Please enter your option :3

1 View ALL apointments
2 View SPL apointments
3 View SFL 1 apointments
4 View SFL 2 apointments
5 View SFL 3 apointments
6 View Junior apointments
7 Return to main menu

Please enter your option :2

SPL
Hearts
Gretna
Edinburgh
2008-01-01
14:00:00
John Binnie
Referee


SPL
Motherwell
Kilmarnock
Motherwell
2008-01-01
14:00:00
Kevin Head
Referee

................................


SPL
Rangers
Falkirk
Glasgow
```

```
2008-01-01
14:00:00
Tom Hanks
Observer
```

The complete set of assignment results can be found in appendix C.

## 4.3 Web interface

A web interface to the Allocator has implemented to enable officials to state their availability and view their assignments. It is acknowledged that the web template is not my own work [10]. However all the functionality provided is my own work.

There is a number of scripting languages e.g. CGI, JSP and PHP. However I think that PHP should be used because it can be embedded into HTML and can run in a web browser and it comes under a GPL compatible license. It can use objects since it is object oriented and it can run under any OS. The way it works is as follows, first it takes the code as input and then it creates web pages as output.

The website has been implemented in PHP and HTML. PHP has been used for the creation of the dynamic content and HTML for the rest part.

Figure 4.3 below illustrates the navigation of the web interface.

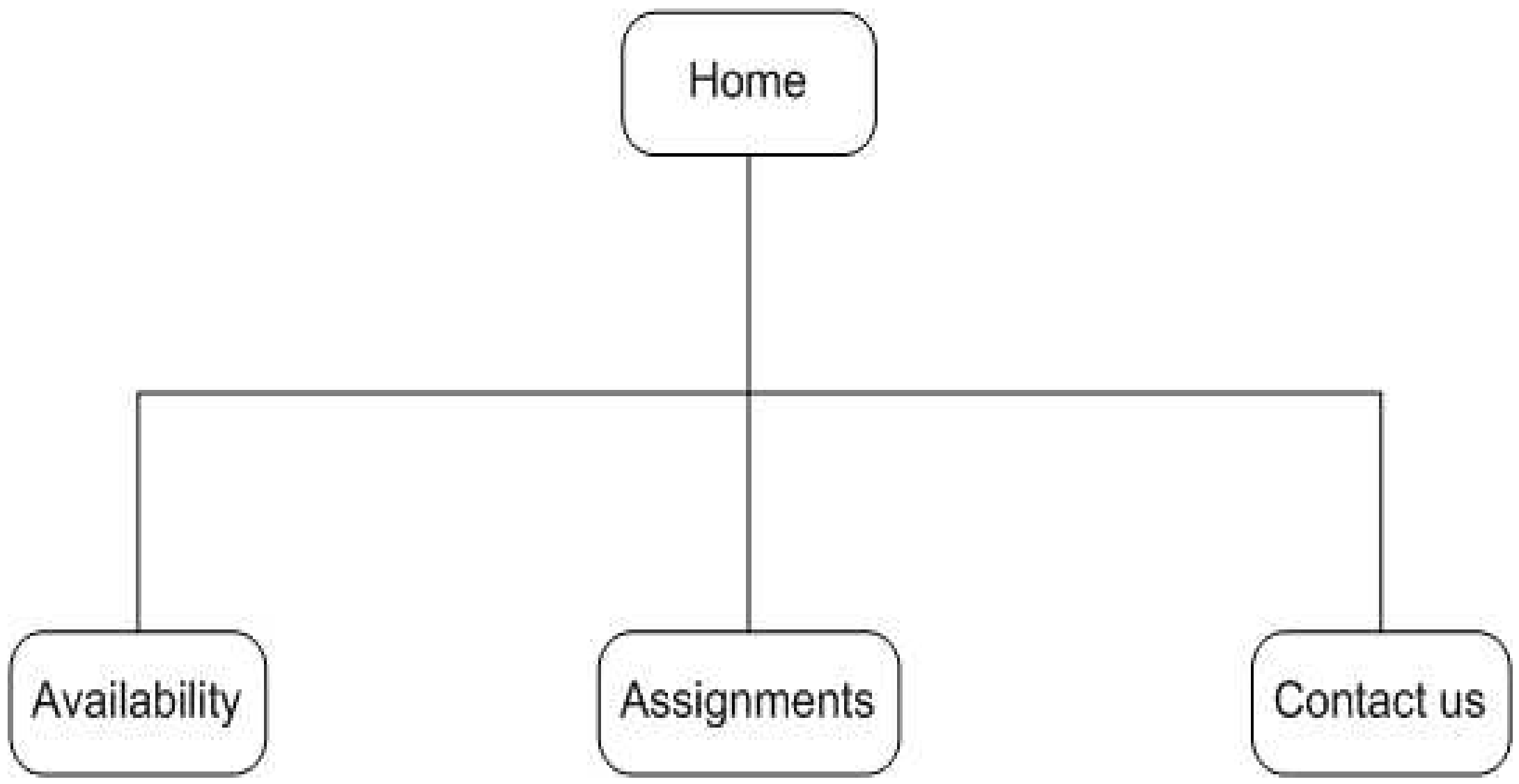

**Figure 4.3** Web Interface Navigation diagram

### 4.3.1 State availability

The web interface gives the opportunity to the official to state his or her availability as shown in figure 4.4.

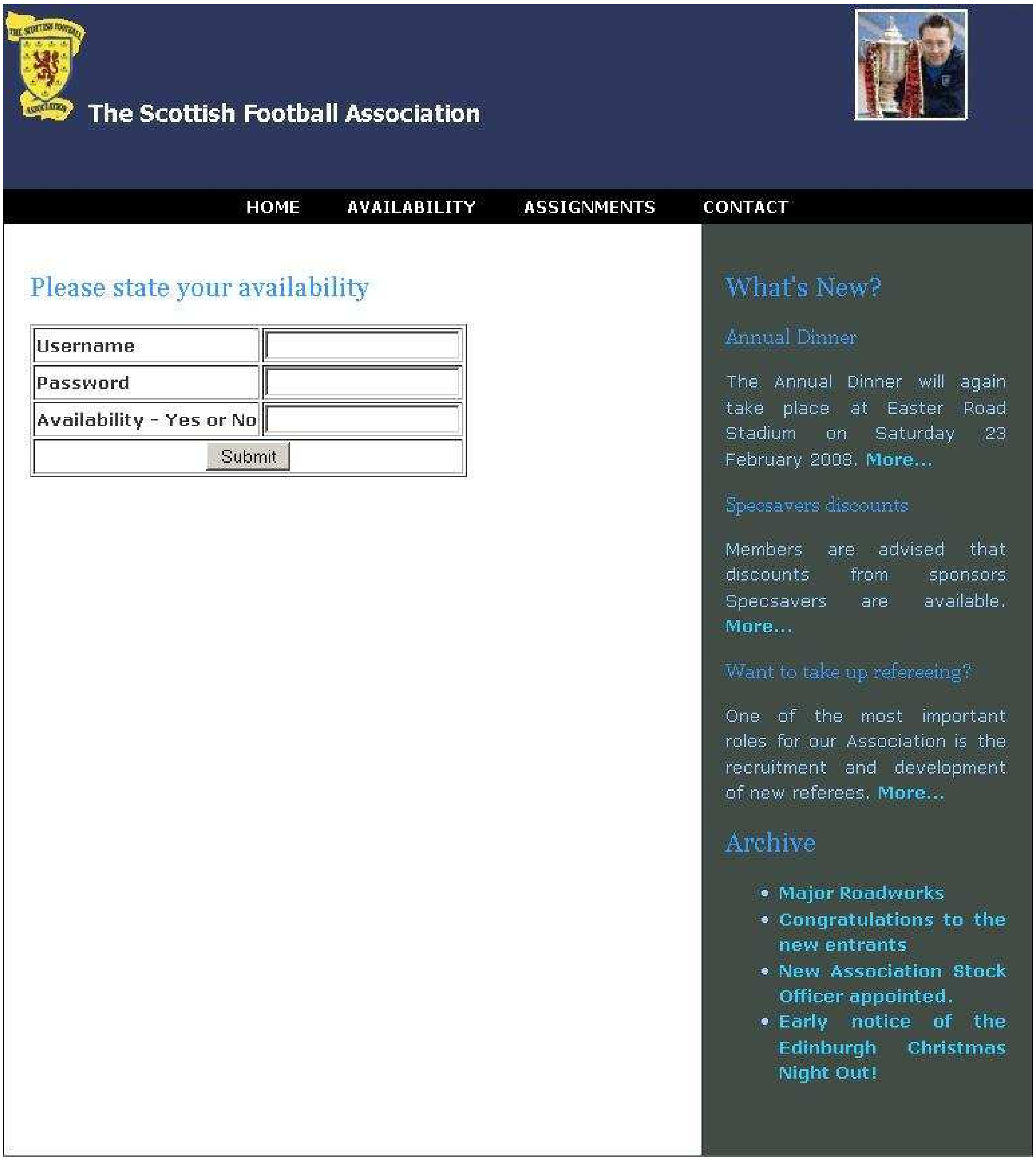

**Figure 4.4** State availability

### 4.3.2 Example assignment

Figure 4.5 below displays an example assignment for Scottish Premier League fixtures, as they would appear on a web browser. The figure it is an image taken from the assignments section of the website. The results are the same as mentioned at section 4.2.5.

| Official's name | Home | Away | Location | Date | Time | Official's role |
|---|---|---|---|---|---|---|
| John Binnie | Hearts | Gretna | Edinburgh | 2008-01-01 | 14:00:00 | Referee |
| Kevin Head | Motherwell | Kilmarnock | Motherwell | 2008-01-01 | 14:00:00 | Referee |
| James Campbell | Rangers | Falkirk | Glasgow | 2008-01-01 | 14:00:00 | Referee |
| Allan Knox | Aberdeen | Celtic | Aberdeen | 2008-01-01 | 14:00:00 | Referee |
| Douglas McDonald | St Mirren | Inverness CT | St Mirren | 2008-01-01 | 14:00:00 | Referee |
| William Lynch | Dundee United | Hibernian | Dundee | 2008-01-01 | 14:00:00 | Referee |
| David Miller | Hearts | Gretna | Edinburgh | 2008-01-01 | 14:00:00 | Assistant Referee 1 |
| Arthur McDonald | Motherwell | Kilmarnock | Motherwell | 2008-01-01 | 14:00:00 | Assistant Referee 1 |
| Paul Hanlon | Rangers | Falkirk | Glasgow | 2008-01-01 | 14:00:00 | Assistant Referee 1 |
| Michael Green | Aberdeen | Celtic | Aberdeen | 2008-01-01 | 14:00:00 | Assistant Referee 1 |
| Alan Mair | St Mirren | Inverness CT | St Mirren | 2008-01-01 | 14:00:00 | Assistant Referee 1 |
| John Thomson | Dundee United | Hibernian | Dundee | 2008-01-01 | 14:00:00 | Assistant Referee 1 |
| Brian Walker | Hearts | Gretna | Edinburgh | 2008-01-01 | 14:00:00 | Assistant Referee 2 |
| Robert Wilson | Motherwell | Kilmarnock | Motherwell | 2008-01-01 | 14:00:00 | Assistant Referee 2 |
| Robert Henderson | Rangers | Falkirk | Glasgow | 2008-01-01 | 14:00:00 | Assistant Referee 2 |
| Calum Bryson | Aberdeen | Celtic | Aberdeen | 2008-01-01 | 14:00:00 | Assistant Referee 2 |
| Guvin Ross | St Mirren | Inverness CT | St Mirren | 2008-01-01 | 14:00:00 | Assistant Referee 2 |
| Sam Holiday | Dundee United | Hibernian | Dundee | 2008-01-01 | 14:00:00 | Assistant Referee 2 |
| Alex Liston | Hearts | Gretna | Edinburgh | 2008-01-01 | 14:00:00 | Fourth Official |
| John Underhill | Motherwell | Kilmarnock | Motherwell | 2008-01-01 | 14:00:00 | Fourth Official |
| Allan Hogg | Rangers | Falkirk | Glasgow | 2008-01-01 | 14:00:00 | Fourth Official |
| Thomas Robertson | Aberdeen | Celtic | Aberdeen | 2008-01-01 | 14:00:00 | Fourth Official |
| Michael Banks | St Mirren | Inverness CT | St Mirren | 2008-01-01 | 14:00:00 | Fourth Official |
| Mark Ireland | Dundee United | Hibernian | Dundee | 2008-01-01 | 14:00:00 | Fourth Official |
| Scott Davidson | Dundee United | Hibernian | Dundee | 2008-01-01 | 14:00:00 | Observer |
| Matthew Perry | St Mirren | Inverness CT | St Mirren | 2008-01-01 | 14:00:00 | Observer |
| Chandler Bing | Aberdeen | Celtic | Aberdeen | 2008-01-01 | 14:00:00 | Observer |
| Matt LeBlanc | Motherwell | Kilmarnock | Motherwell | 2008-01-01 | 14:00:00 | Observer |
| Joy Tribianni | Hearts | Gretna | Edinburgh | 2008-01-01 | 14:00:00 | Observer |
| Tom Hanks | Rangers | Falkirk | Glasgow | 2008-01-01 | 14:00:00 | Observer |

**Figure 4.5** An assignment of officials for SPL fixtures

## 4.4 Evaluation Conclusion

The feedback received from two league secretaries, after they used the Allocator was positive in general. However they both said that it is fairly important to have the option to use the software for multiple weeks. Also the manual assignment it would be useful to have if supplied since the change assignment option is usually used if someone calls of a game and they have to assign someone else. They weren’t very interested on the web interface because a significant number of officials are not competent enough with the internet. However they used it in order to complete the evaluation. The comments were positive and the only thing that should be changed is the availability section, which should give the option to the officials to state their availability for a certain date.

# Chapter 5

# Final System Design and Implementation

This chapter outlines the design and implementation of the final system. The Allocator comprises a database, an allocation algorithm and a web interface. The design can be found in section 5.1, the implementation in section 5.2 and the web interface in section 5.3. The source code for the final system can be found in appendix A.

## 5.1 Design

The final extends the prototype with modification to assign officials manually and for multiple weeks. The Web interface is based on the prototype version with a modification to state availability for a certain date. These modifications completed after feedback received from SFA secretaries during a demonstration of the prototype as described in section 4.4. The algorithm used still remains the same as the prototype and its description can be found in section 4.1.2.2.

### 5.1.1 Database Design

The design of the database aims to keep it as simple and efficient as possible. The official table will hold the official's name, id, category, experience, username and password. The availability table will hold the official's id, which will be referenced from the official table and a date. The fixture table will hold the fixture id, the type of the fixture, the home and away teams, the location, the date and the time. The assignment table will hold the official's id which will be referenced from the official table, the fixture id which will be referenced from the fixture table and the official's role.

The database extends the database of the prototype system and the technologies described at section 4.1 have been used.

Figure 5.1 below shows the Entity Relationship diagram designed for the database of the final system.

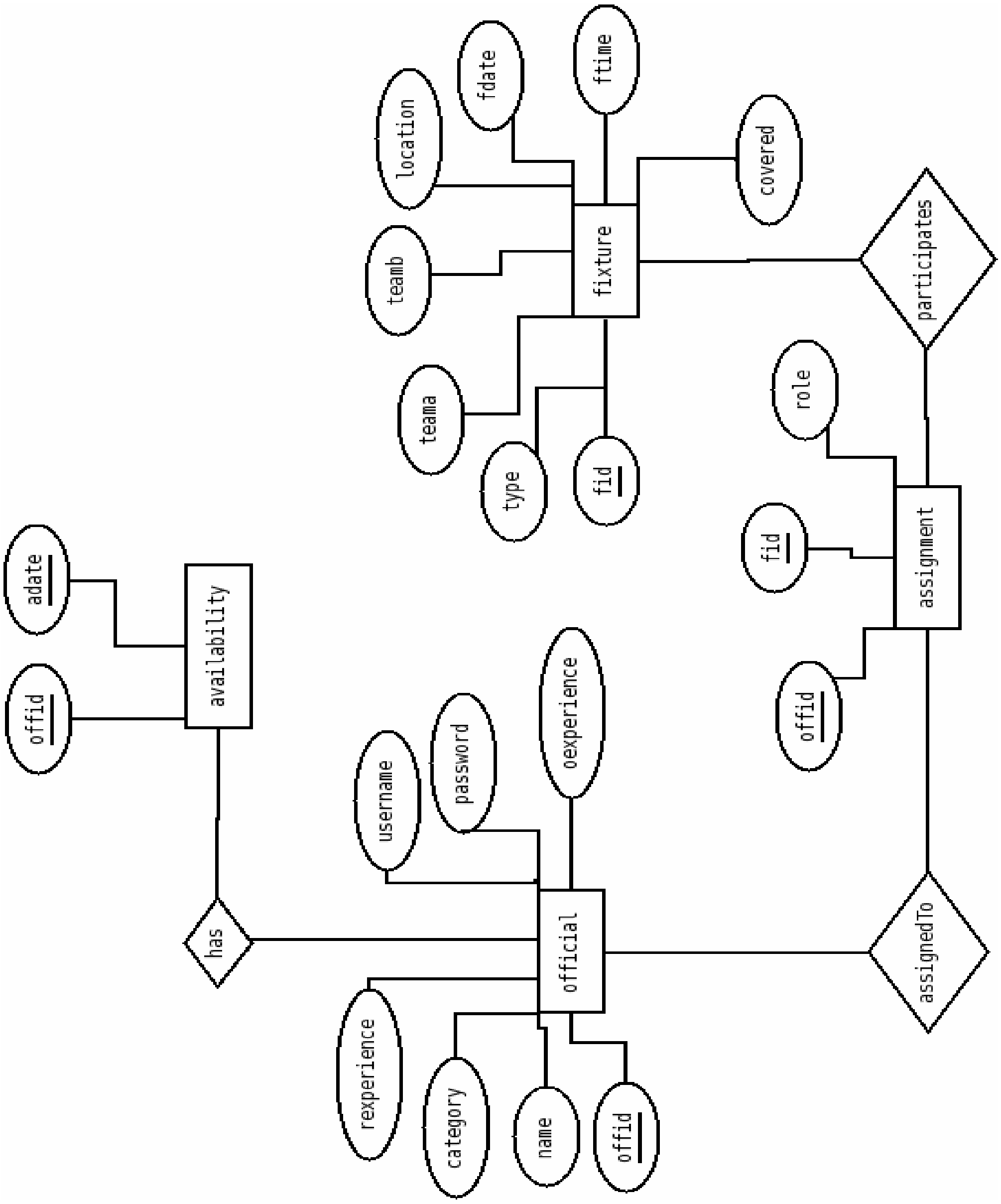

**Figure 5.1** SFA Allocator Entity Relationship Diagram

### 5.1.2 Allocator Design

A final usable referee allocation system has been designed that uses the greedy algorithm, as described in section 4.1.2.2, and is extended after evaluation of the prototype. The final system extends the prototype with modification to assign officials manually or by combining manual and automatic options. It also gives the option to assign officials for multiple weeks.

### 5.1.3 Web Interface Design

The Web interface is based on the prototype, as described in section 4.3. Section 5.3 describes the final version of the Web Interface after the evaluation of the prototype by SFA secretaries.

## 5.2 Implementation

The final system includes a command line user interface with the following options: Update officials and fixtures, Assign officials, View appointments, change assignment and exit. All these properties work and have sub properties attached to them. The final system aims to give the option to add or remove officials and fixtures, assign officials to fixtures and view the appointments and manually assign officials.

Table 5.1 below shows the lines of code for each part of the system. The technologies used are described in chapter 2. Also the source code of the final system can be found in appendix A.

| Technology | Lines of Code |
|---|---|
| Java | 817 |
| PHP | 100 |
| SQL | 40 |

**Table 5.1** Lines of Code

Figure 5.2 below is a diagrammatic representation of the Allocator main menu.

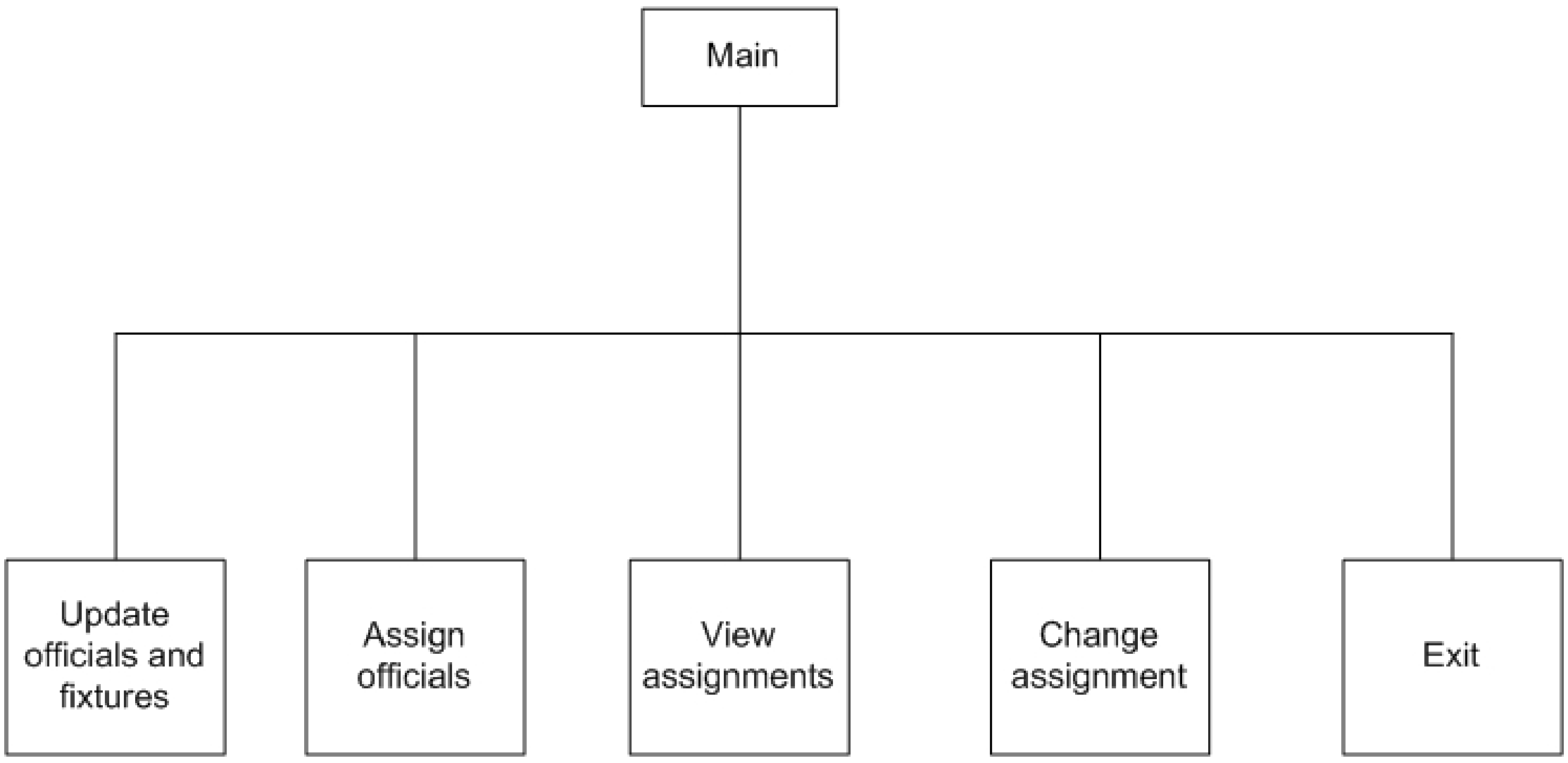

**Figure 5.2** Diagrammatic representation of the Allocator main menu

### 5.2.1 Update officials and fixtures

The update officials and fixtures option of the menu takes us to a new menu with the options to add or remove officials and fixtures and return to the main menu.

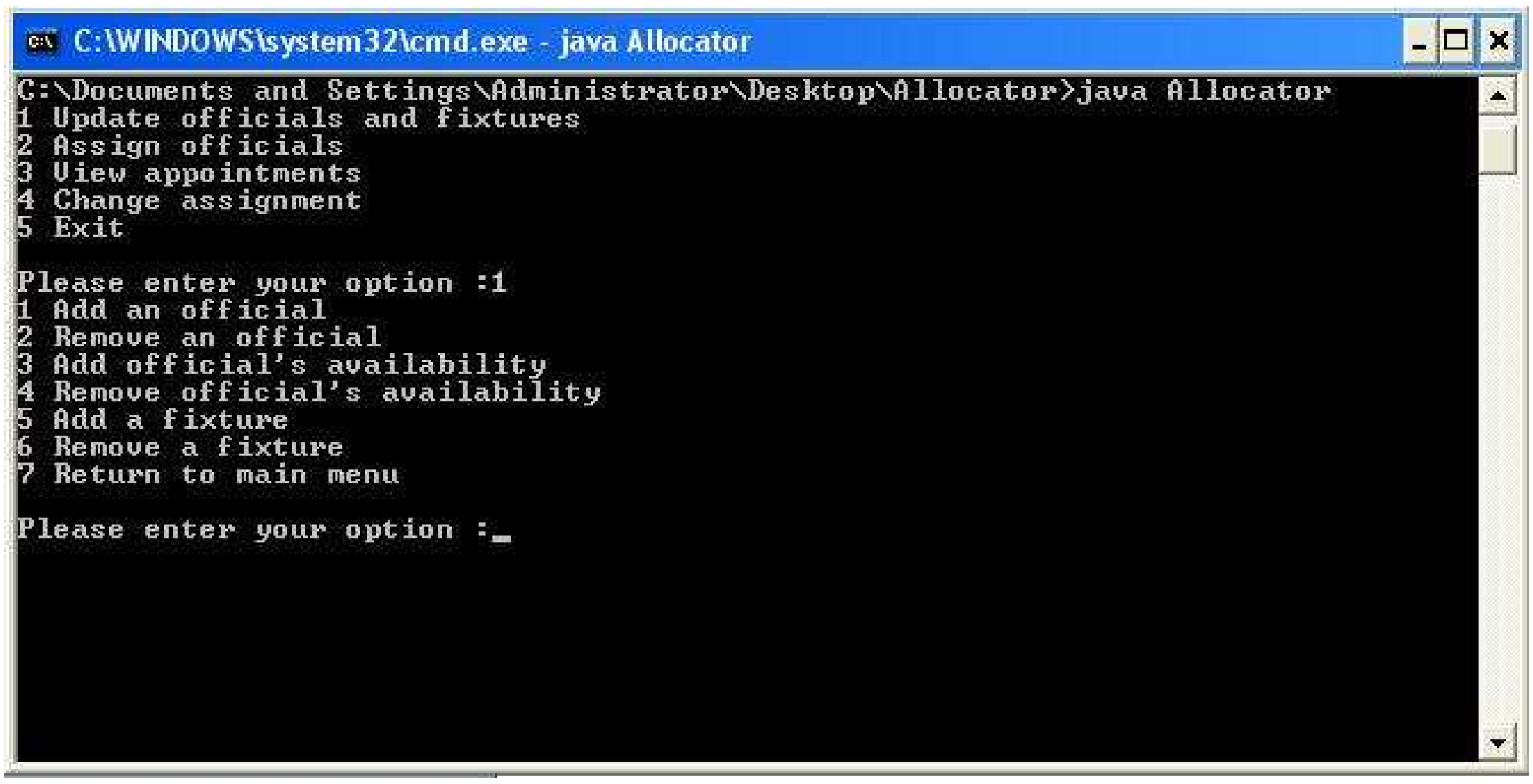

**Figure 5.3.1** Update Officials and Fixtures

### 5.2.2 Assign officials

The assign officials, option takes us to a new menu with the options to assign officials at Scottish premier league, Scottish football league 1, Scottish football league 2, Scottish football league 3 and junior football. The assignment can be done separately for every league or with the assign all option which will assign officials to every fixture.

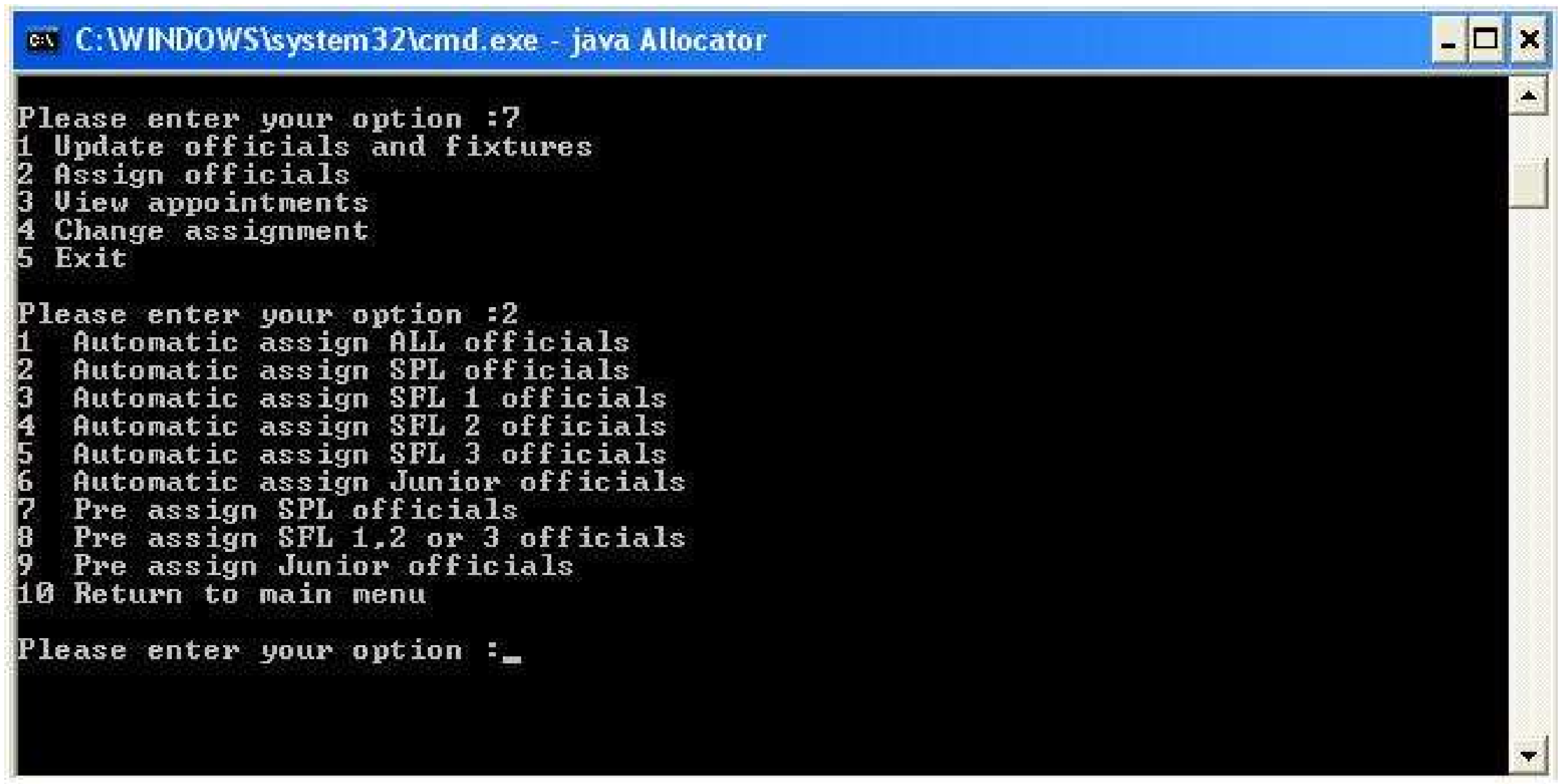

**Figure 5.3.2** Assign Officials

### 5.2.3 View Appointments

The view appointments option gives us the opportunity to display on the screen the appointments to fixtures. This can be done for every league separately or for all leagues and for a certain date.

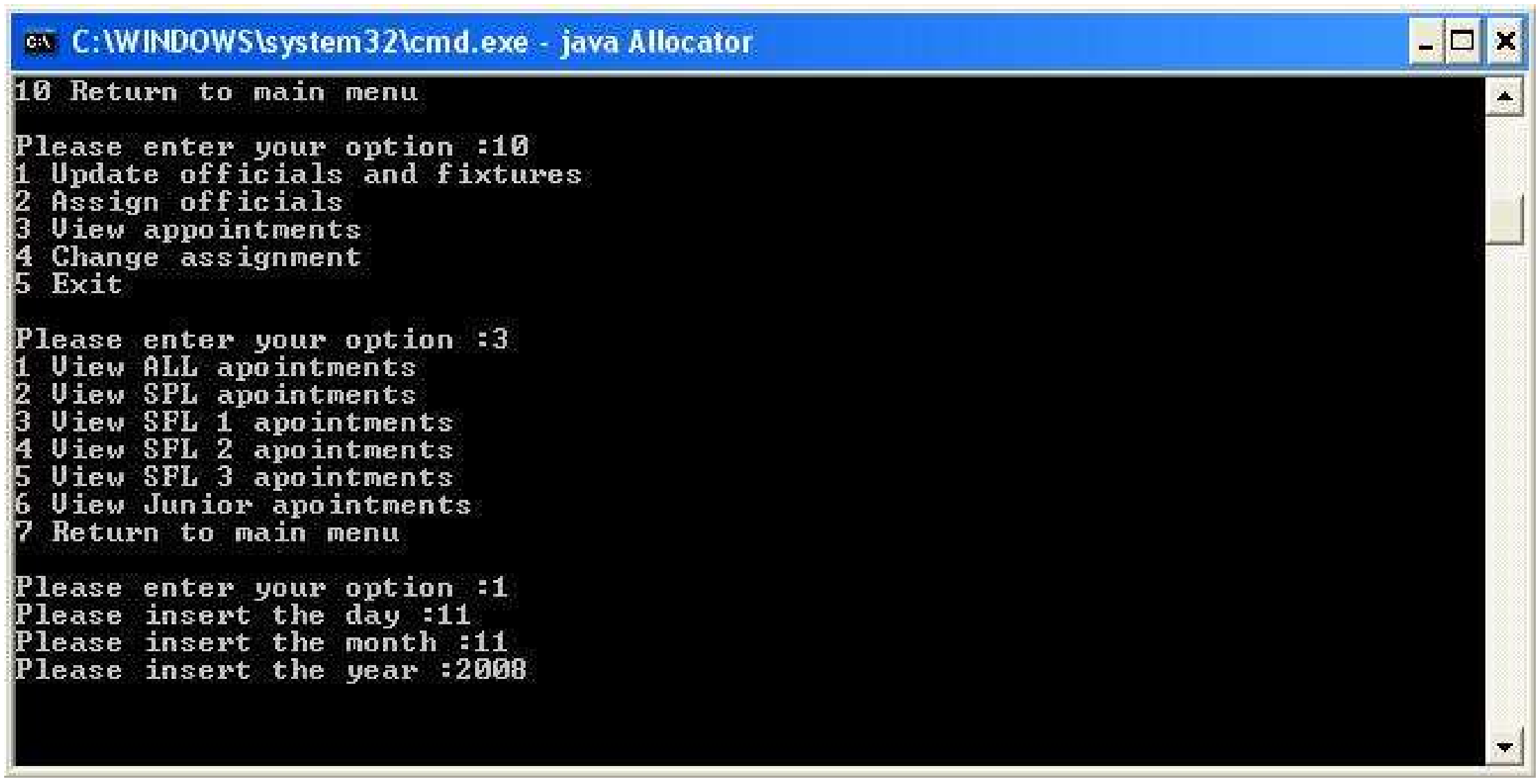

**Figure 5.3.3** View Assignments

### 5.2.4 Change assignment

The option to change an assignment is also given.

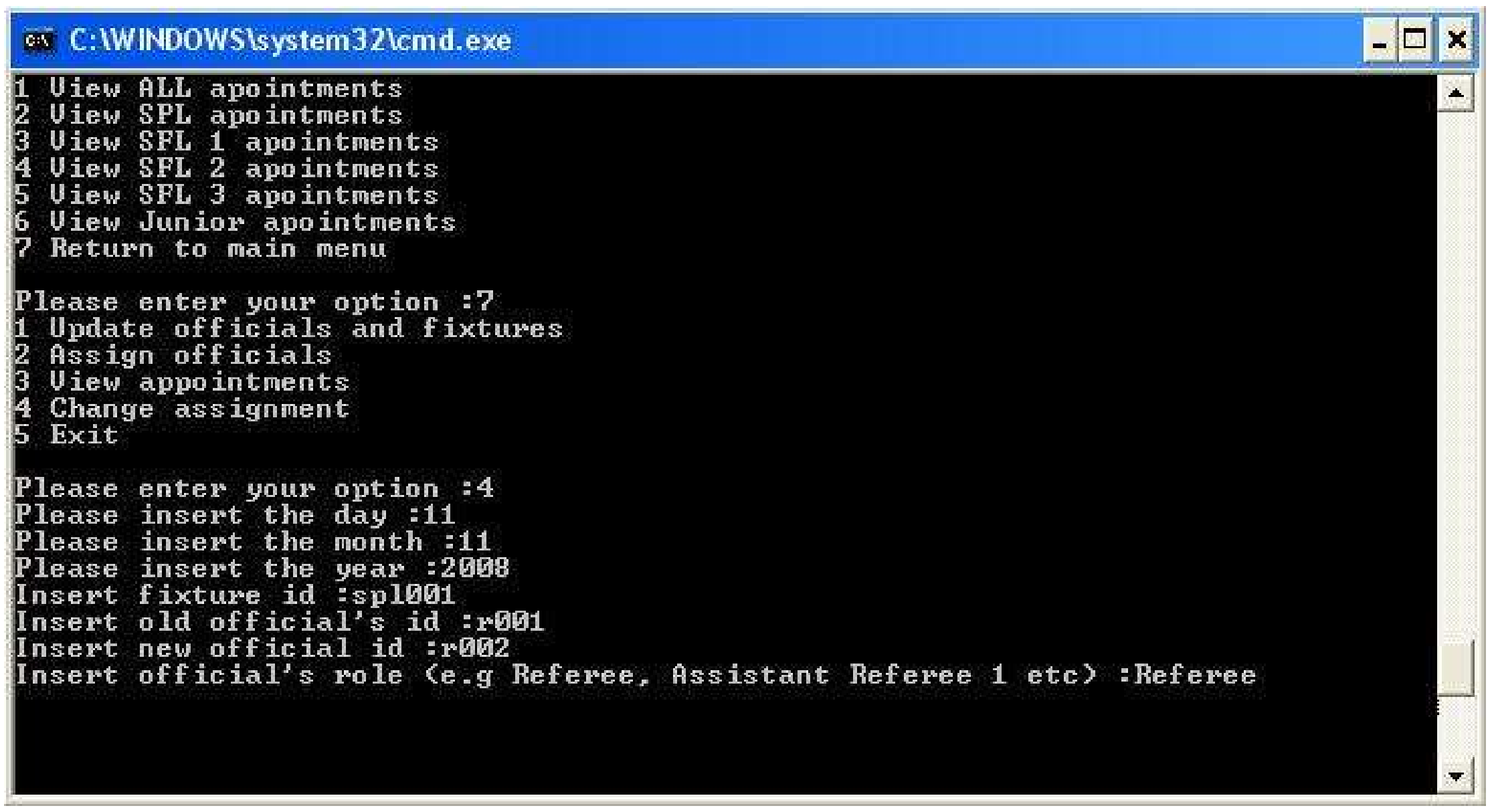

**Figure 5.3.4** Change an Assignment

### 5.2.5 Manual assignment

Finally the option to assign a number of officials manually and the rest automatically has been given.

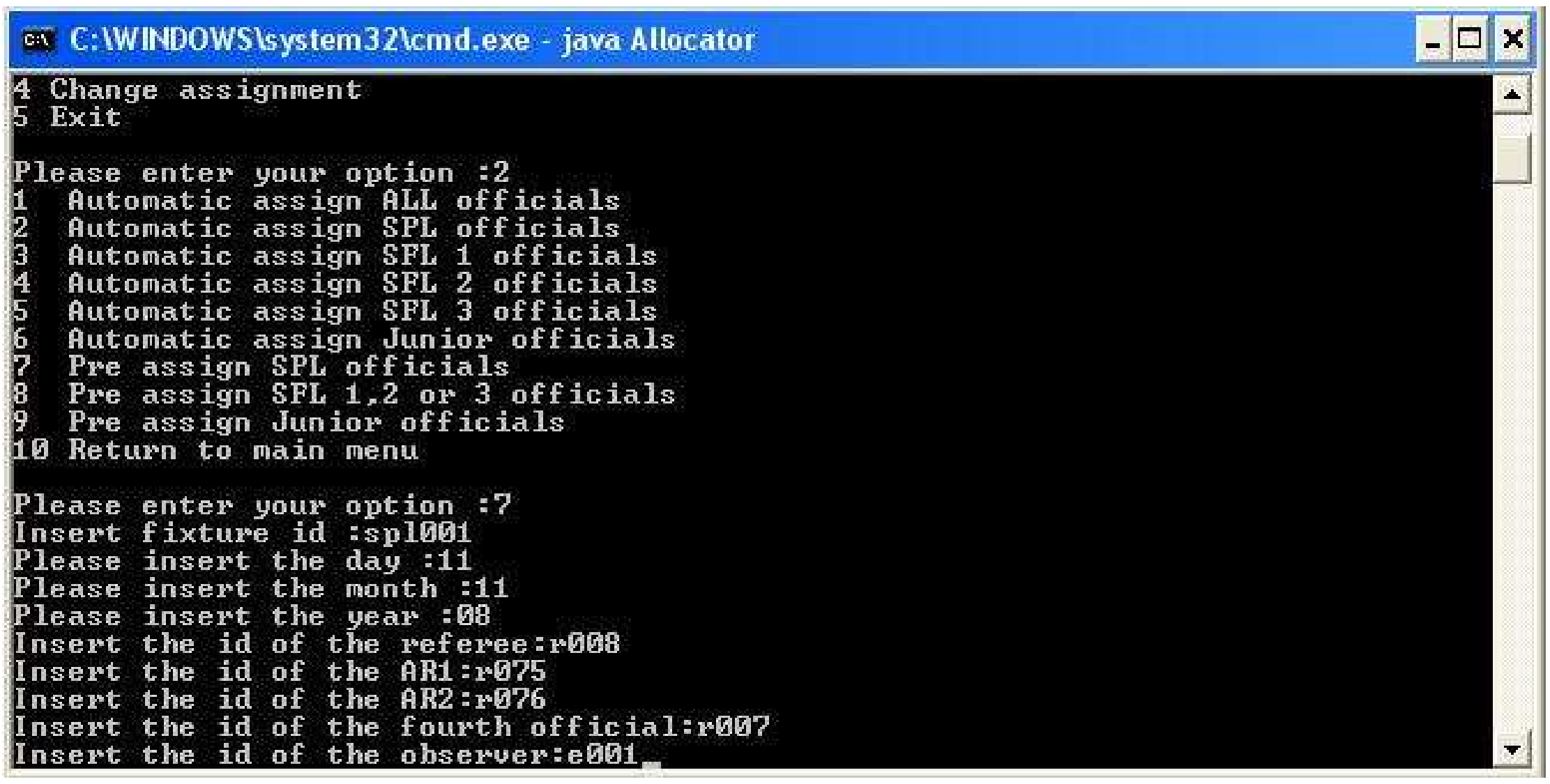

**Figure 5.3.5** Manual Assignment

### 5.2.6 Multiple weeks

The option to assign officials for multiple weeks is available.

### 5.2.7 Modeling

A number of UML class diagrams are used to describe the final system.

Figure 5.4 shows the final version of the main class of the Allocator. In this case the class contains methods to retrieve officials and fixtures and methods to assign officials to fixtures, change assignment and view the assignments.

| **Allocator** |
|---|
| fixtureNo: int<br>max: int<br>premierReferee1[]: official<br>premierFixtures[]: fixture<br>fi: String<br>rid: String<br>ar1id: String<br>ar2id: String<br>fourthid: String<br>obid: String<br>giveDate: String<br>driverName: String<br>dburl: String<br>con : Connection |
| +mainMenu() : void<br>+updateMenu() : void<br>+assignMenu() : void<br>+viewMenu() : void<br>+addOfficial() : void<br>+removeOfficial() : void<br>+addAvailability() : void<br>+removeAvailability() : void<br>+addFixture() : void<br>+removeFixture() : void<br>+askForDate() : void<br>+preassignSPL() : void<br>+ preassignSFL() : void<br>+askforfixture() : void<br>+askforreferee() : void<br>+askforar1() : void<br>+askforar2() : void<br>+askforfourth() : void |
| +askforobserver() : void<br>+preassignreferee() : void<br>+preassignar1() : void<br>+preassignar2() : void<br>+preassignfourth() : void<br>+preassignobserver() : void<br>+updatefixtureset() : void<br>+getRef() : void<br>+getObserver() : void<br>+getFixture() : void<br>+getAssistant() : void<br>+assign() : void<br>+assignSPL() : void<br>+assignSFL1() : void<br>+assignSFL2() : void<br>+assignSFL3() : void<br>+assignJUNIORS() : void<br>+view() : void<br>+changeAssignment() : void |

**Figure 5.4** UML Class diagram for the Allocator

Figure 5.5 shows the final version of the official object, which is used in the main to temporary hold officials before assigned.

| **Official** |
|---|
| offid : String<br>name : String<br>category : int |
| +official :<br>+official :<br>+official : |

**Figure 5.5** UML Class diagram for the official object

Figure 5.6 shows the final version of the fixture object, which is used in the main to temporary hold fixtures before officials are assigned to them.

| **Fixture** |
|---|
| fid : String<br>type : String<br>teama : String<br>teamb : String<br>location : String |
| +fixture :<br>+fixture : |

**Figure 5.6** UML Class diagram for the fixture object

The use case diagram describes the functionality of the final system as designed from the final requirements as described in chapter 3.

## 5.3 Web interface

The web site will include a main page, a page to state availability for a certain date, a page to view the assignments and a contact us page. The website is the same as prototype as described in section 4.3. The only difference is that the officials can state their availability for a certain date as shown below in figure 5.7.

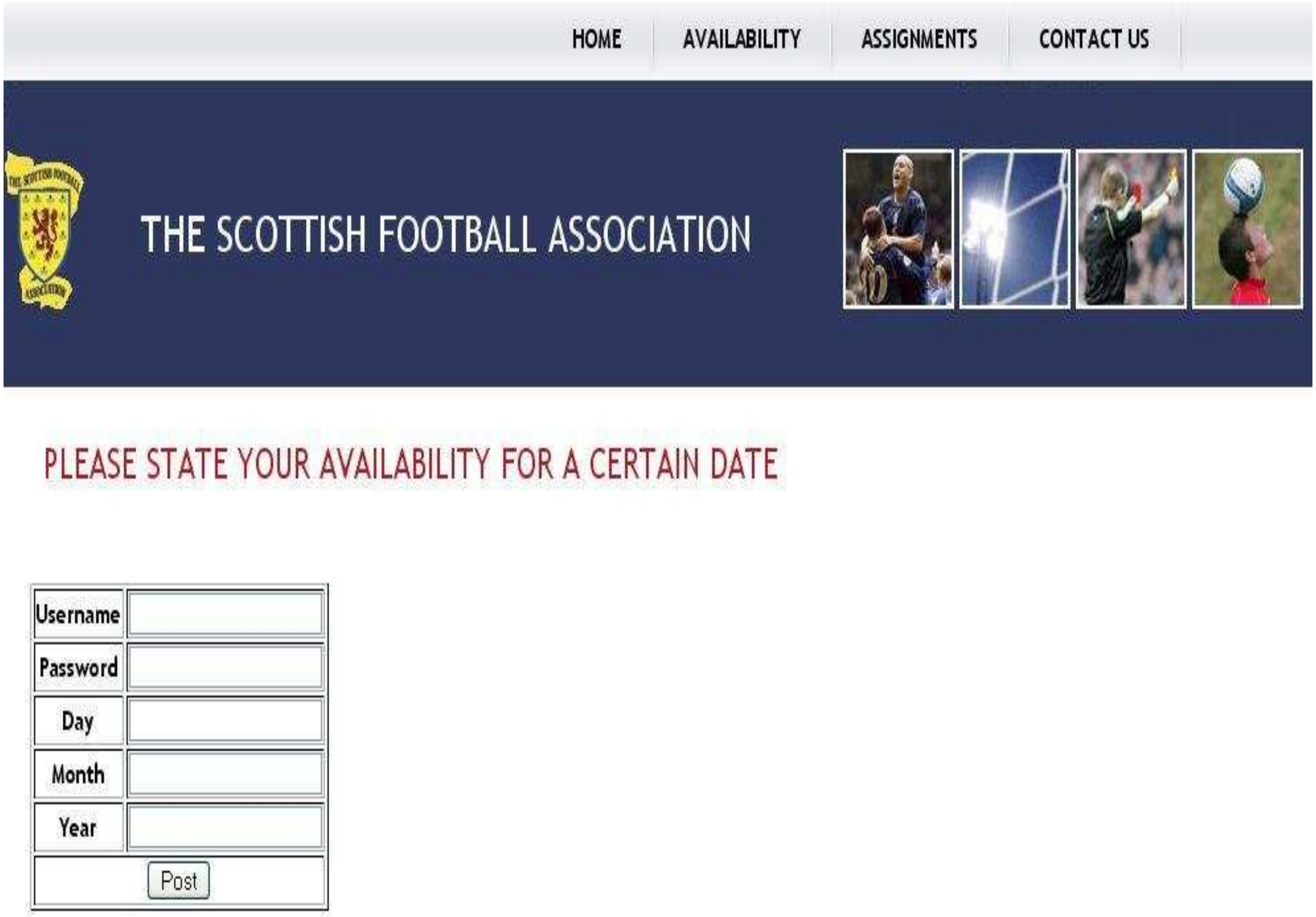

**Figure 5.7** State Availability

# Chapter 6

# Testing and Evaluation

---

This chapter discusses the testing and evaluation of the final system. The testing strategy used can be found in section 6.1, the evaluation in section 6.2 and the feedback received from the secretaries that used the final system in section 6.3. The testing tables can be found in appendix E.

## 6.1 Testing

A testing strategy knows as Unit and Integration testing [11] has been used to check that the system behaves as expected. The testing strategy was based on the functionality and the requirements of the system. Separate tests took place for both the Command Line and the Web Interface. The testing tables can be found in appendix E.

### 6.1.1 Command Line system testing

A unit test was performed for every functional component as reported in appendix E. The system can successfully update officials and fixtures, assign officials, vies appointments, change assignments.

### 6.1.2 Web Interface testing

A unit test was performed for every functional component as reported in appendix E. The officials can use the web interface to state their availability and view their appointments. The Web interface was also tested under a variety of different system settings, operating systems and browsers to ensure that is usable.

## 6.2 Evaluation

A Major part of this project was to build a usable system, based on real world requirements and to evaluate it using real data sets by SFA secretaries. Looking back at the requirements section, the final system managed to satisfy every major requirement, including new requirements introduced after feedback received from SFA secretaries. However due to time constraints it wasn't possible to develop the statistics part of the project, which was optional.

### 6.2.1 Command Line system evaluation

The Command Line system was evaluated by two SFA secretaries to assign officials. The feedback received from them can be found in section 6.3. Also the prototype system was evaluated by two SFA secretaries as described in section 4.4.

### 6.2.2 Web Interface evaluation

The Web Interface was evaluated by two SFA officials and was found sufficient enough, as described earlier in section 4.4. However due to the internet illiteracy of a significant amount of officials they do not wish to use professionally the Web Interface.

### 6.2.3 Performance

The response time of the algorithm is vital and it should be kept at the lowest level possible. The greedy method took about five seconds to assign approximately 100 officials using a 2.1 GHz computer with 512mb of ram running Windows XP for a certain date. The table below show the average and total number of fixtures that are usually set up to play on a football day.

| Type | No Of Fixtures | Officials Required |
|---|---|---|
| SPL | 6 | 30 |
| SFL1 | 5 | 20 |
| SFL2 | 5 | 20 |
| SFL3 | 5 | 20 |
| Junior Football | 5 | 20 |
| **Total** | 26 | 110 |

**Table 6.1** An assignment of officials using the Greedy method

## 6.3 Feedback from secretaries

The feedback received from two league secretaries, after using the final system, was positive in general. However they both said that it would be useful to have the option to assign the officials using their name and not their id. The secretaries also stated that they did not find any difficulties while using the Command Line system and that the navigation of the menu is straight forward. After the requirements set by them, after the demonstration of the prototype have been met they said that they can now use the system using real data sets. The letter of the secretaries can be found in appendix F.

# Chapter 7

# Discussion

This chapter begins by summarizing the project in section 7.1 then it goes on describing the main achievements in 7.2 and limitations in 7.3. Future work is discussed in section 7.4. The methodology used is described in section 1.4.

## 7.1 Summary

The main objectives of the project were:

To develop an Allocation algorithm that will be used to assign officials to different levels of football. The algorithm can be found in section 4.1.2.2. To develop a database were all the information about the officials, the fixtures and the assignments will be stored. To develop a Web Interface to the Allocator were the officials will be able to state their availability and view the assignments. To develop a website displaying statistical information about the officials, such as the correctness of decisions, the match control, the management of players and team officials and the average number of sanctions. This part was optional.

The following objectives were subsequently met:

A background research took place, which included an overview of the referee assignment in Scotland and any relevant referee assignment tools available. A database was developed, which stores information about the fixtures, the officials and the assignments. A usable prototype system was designed and developed, which using a greedy algorithm, can assign officials to different levels of football. A Web Interface to the system was developed. The officials can use this website to state their availability and view their assignments. The prototype was evaluated by SFA secretaries and feedback received and new requirements introduced after the evaluation of the prototype. A final Command Line system that extended the prototype was developed which satisfied all of the main requirements. The Web Interface was updated to meet the new requirements. As a final point I would like to state that the final system was used by SFA secretaries to assign officials.

## 7.2 Main achievements

The project has met or exceeded almost all of the objectives. Looking back at the requirements section in chapter 3, the final system managed to satisfy every major requirement, including new requirements introduced after feedback received from SFA secretaries. However due to time constraints it wasn’t possible to develop the statistics part of the project. The letters of the secretaries who evaluated the system can be found in appendix F.

The following achievements were subsequently met. Investigated a range of algorithms for solving the referee allocation problem, including backtracking/constraint solving and developed a greedy allocation algorithm. A usable referee allocation system was designed that is based on the greedy algorithm, which has been extended after consultation with the SFA. Implemented a prototype system and evaluated it with SFA secretaries. Implemented a final system based on the prototype, with both a command line and a web interface. The final system was evaluated by SFA secretaries. The system has been used by SFA secretaries to assign officials and it will be used again in the future. Their letters of recommendation can be found in Appendix F. Completed a literature research within the SFA to learn about the referee allocation procedure in general, established the requirements with consultation with the SFA and investigated existing tools.

## 7.3 Limitations

Despite meeting all of the main requirements, there are some limitations to the system.

### 7.3.1 Command Line system limitations

The system does not contain a search function for retrieving official and fixture details and does not give the option to edit the details of officials and fixtures. Also every time we wish to assign an official manually we must use his id, rather than his name.

### 7.3.2 Web Interface limitations

A general log in option is not available. The website does not contain a search function and the options to view assignment for a certain date only of for a certain official are not available.

## 7.4 Future Work

There are a number of possible extensions which could be added to the work it the time was sufficient.

### 7.4.1 Command Line system future work

The SFA secretaries expressed a desire to have the ability to assign officials using their name and not their id. One way to do this is to change slightly the database and the parts of the code asking for the id with the name. The only issue with this is when two or more officials have the same name, which could be solved by asking who we want.

It would be advantageous to have a search facility to the system. Testing revealed that it would be useful to be able to see how many fixtures there are no a certain date, before we assign officials to them and see the details of the officials.

Last but not least would be the option to edit officials and fixtures. At the moment we can only add officials and fixtures. It would save us time if we could edit a fixture or an official and change an attribute of it, instead of deleting it and adding it again.

### 7.4.2 Web Interface future work

Regarding the Web Interface it would be beneficial to have a statistics part, which will display information about different aspects of refereeing such as the correctness of decisions, match control, management of players and team officials and the average number of sanctions.

It would be helpful to give the ability to the officials to log in to the site if they wish to state their availability. At the moment they have to enter their user name, password and date every time they wish to state their availability for a certain date.

# References

1. Sharp, Rodgers, Preece 2007 *Interaction Design 2nd* Ed.
   Wiley p530

2. Jeff Wigal *About Referee Assistant*. Available from:
   http://www.referee-assistant.com/about/
   [Accessed 29 May 2008]

3. Shana *Saavedra Soccer Central Software*. Available from:
   http://www.yoursoccercentral.com/frameset.php
   [Accessed 30 May 2008]

4. Sun Microsystems *The Java Technology.* Available from:
   http://java.sun.com/docs/books/tutorial/getStarted/intro/definition.html
   [Accessed 08 May 2008]

5. Three Tier Architecture
   http://java.sun.com/docs/books/tutorial/getStarted/intro/definition.html
   [Accessed 20 May 2008]

6. Relational Database Management Systems
   http://aspalliance.com/1211_Relational_Database_Management_Systems__Concepts_and_Terminologies.all
   [Accessed 23 May 2008]

7. Comparing Web Scripting and Website Programming Languages
   http://training.gbdirect.co.uk/courses/php/comparison_php_versus_perl_vs_asp_jsp_vs_vbscript_web_scripting.html
   [Accessed 23 May 2008]

8. Sartaj Sahni 2000 *Data Structures, Algorithms and Applications in Java*.
   P834-963

9. Sartaj Sahni 2000 *Data Structures, Algorithms and Applications in Java*.
   P699-742

10. Reality Software and Web solutions
    http://www.realitysoftware.ca/
    [Accessed 28 May 2008]

11. Ron Patton 2005 *Software Testing 2nd* Ed.
    Sams